\documentclass[preprint,12pt]{article}
\usepackage{authblk}
\usepackage{braket}
\usepackage[square, sort&compress, numbers]{natbib}
\usepackage{color}
\usepackage{amssymb,amsmath,amsmath,amsfonts,amsthm,bm}
\usepackage{hyperref}
\usepackage{xcolor}
\hypersetup{
    colorlinks,
    linkcolor={red!75!black},
    citecolor={blue!75!black},
    urlcolor={blue!75!black}}
\usepackage[utf8]{inputenc}
\usepackage{physics}
\usepackage{graphicx}
\usepackage{caption}
\usepackage{subcaption}
\usepackage{tabularx}
\usepackage{enumitem}
\usepackage{cleveref}
\usepackage{threeparttable}
\usepackage{lscape}
\usepackage{natbib}
\usepackage{booktabs}
\graphicspath{{images/}}
\usepackage{cancel}
\usepackage[normalem]{ulem}
\usepackage{hyperref}      
\usepackage{flushend}
\usepackage{color}
\usepackage{float}
\usepackage{multirow}

\definecolor{red}{rgb}{1.0,0.0,0.0}

\begin{document}
\title{Thermal conductivity of dense $np\Lambda$ matter in neutron star cores}

\author{Tanay Choudhary$^{1}$,
Athira S.$^{1}$,
  Monika Sinha$^{1}$\thanks{Corresponding author: ms@iitj.ac.in}\\
  {\small $^{1}$Indian Institute of Technology Jodhpur, Jodhpur 342037, India}}
\maketitle

\begin{abstract}
The possible presence of hyperons in the cores of massive neutron stars has important implications for their microscopic transport properties and thermal evolution. Despite recent progress in modelling core transport, a dedicated analysis regarding the thermal conductivity of a specific \(np\Lambda\) mixture remains absent from the existing literature. In this work, we investigate the thermal conductivity of dense, \(\beta \)-equilibrated \(np\Lambda\) matter within the framework of the variational linearized Boltzmann kinetic approach, employing the density-dependent DDME2 equation of state across a baryon density range of \((0.5\text{--}4.5)\,n_0\). We find that neutrons still dominate thermal transport, while the onset of \(\Lambda \) hyperons induces only a remarkably small reduction in neutron conductivity compared to pure nucleonic matter. These results suggests that the core thermal relaxation timescale remains practically unaltered in the presence of $\Lambda$ hyperons.
\end{abstract}

\maketitle


\section{Introduction}
\label{intro}

The unique stellar objects, neutron stars (NS), are born in supernovae with high temperature $\sim 10^{11}$ K \cite{Yakovlev_2004, RevModPhys.64.1133}. However, within a few days, its surface temperature goes down to $\sim 10^9$ K, while the inner region of the star remains still hotter compared to its surface. Then, with time, the crust and the core of the star cool down independently with different mechanisms. In this stage, the thermal equilibrium between the different parts of the star is not achieved because of the insufficient thermal conduction in the interior of the star \cite{Sales_2020}. The core of the star cools down rapidly by neutrino emission while the crust remains hotter \cite{Page_2006,Yakovlev_2001}. This results in heat flow from crust to the core within a short period of time of $10-100$ years. This duration is known as the thermal relaxation era \cite{Gnedin_2001}.

Heat transport in the dense interiors of NS is essential for interpreting their thermal evolution and observable surface temperatures. It plays a crucial role throughout a neutron star; however, its significance is particularly pronounced in the core, where the bulk of the stellar mass and thermal energy reside. Therefore, the study of the thermal conductivity of dense matter in the neutron star interior is crucial for understanding heat transport \cite{Potekhin_2015,Shternin_2022,Baiko_2001}. The core is also the primary site of neutrino emission \cite{Yakovlev_2001}, which dominates the cooling of neutron stars over most of their evolution. In this context, thermal conductivity governs the efficiency of heat redistribution, thereby directly influencing the global cooling behavior \cite{PhysRevD.75.103004, Yakovlev_2005}. In contrast, the crust contains only a small fraction of the total heat capacity and contributes less significantly to the overall energy budget \cite{PhysRevC.95.025806,page2012thermaltransportpropertiesneutron, Burrello2015}. While transport properties in the crust regulate the flow of heat to the surface and thus affect observable temperatures, their impact is largely confined to transient phenomena, such as the thermal relaxation of young neutron stars or the surface response in accreting systems \cite{Yakovlev_2004c}. Therefore, the thermal conductivity of the core is of central importance for modelling neutron stars' cooling, as it determines the internal thermal structure, controls the efficiency of heat transport, and plays a key role in connecting observational signatures to the underlying properties of dense matter \cite{Tsuruta_2009,Potekhin_2015}.

Recent astrophysical observations suggest the existence of massive NS with the mass above $2~M_\odot$ \cite{Demorest_2010, Antoniadis:2013pzd, NANOGrav:2019jur,Zdunik_2013}. The existence of massive NS opens up the possibility of the appearance of high-mass particles like hyperons -- the massive strange baryons inside the inner core of the NS leading to the formation of a hyperon star \cite{Oertel_2016, GlendenningHyperons}.

The thermal conductivity of highly dense matter depends on the composition of the matter \cite{Baiko_2001, Gnedin_1995, ShterninVidana2021}. While significant progress has been made in quantifying the thermal conductivity of nucleonic matter, particularly neutrons, the role of strange baryons in this context remains comparatively unexplored (see, for example, chapter 9 of Ref.\cite{RezzollaEtAl2017} for a review of transport phenomena in neutron stars). 

Early investigations of thermal conductivity of baryons in NS cores have primarily focused on nucleonic components, with seminal contributions providing a detailed microscopic description of neutron thermal conductivity \cite{PhysRevB.35.1620, PhysRev.185.323, Flowers1979}.
The microscopic calculation of neutron thermal conductivity in neutron star cores ($\rho \sim (1\text{--}8)\times 10^{14}\ \mathrm{g\,cm^{-3}}$) was evaluated in detail by Baiko, Haensel, and Yakovlev \cite{Baiko_2001}. They utilized nucleon-nucleon transition probabilities derived from Dirac–Brueckner many-body approaches \cite{Li_1992,Machleidt_1989,Brockmann_1990} with the Bonn potential \cite{Machleidt_1987,Machleidt_1989} while comprehensively accounting for neutron–neutron and neutron–proton collisions as well as superfluid effects.

However, at the high densities characteristic of massive neutron star interiors, the composition of matter is expected to extend beyond nucleons to include strange baryons \cite{Glendenning:1984jr,Logoteta2021,Bednarek_2012}. In particular, $\Lambda$ hyperons are among the first exotic species to appear and can significantly influence both the equation of state (EOS) and transport properties \cite{Negreiros_2018,Gusakov_2014, Raduta_2018,Banik_2014}. These additional degrees of freedom introduce new scattering channels and modify existing ones, potentially leading to nontrivial changes in transport coefficients. Despite this, the thermal conductivity associated with hyperonic degrees of freedom has received comparatively little attention, and a systematic treatment of $\Lambda$-mediated heat transport is still lacking in the literature. This gap is driven by a severe lack of experimental data in the hyperonic sector. This is particularly pronounced for hyperon-hyperon interactions, where empirical constraints are virtually nonexistent; indeed, the foundational data available relies almost entirely on three double-hypernuclei emulsion events recorded prior to 2001 \cite{gal2016strangeness}. This causes the study of transport coefficients, where the experimental scattering data is given as input, to be heavily model-dependent and therefore highly uncertain. Nevertheless, with the upcoming proposed experiments for the hyperon scatterings, such as  J-PARC E40 \cite{Miwa:2022coz} and BESIII \cite{miao2024hyperonnucleusnucleonscatteringbesiii}, models can be more constrained, and we can hope to have more precise results for transport coefficients of hyper-nuclear matter.

The most detailed study on the transport properties of the neutron star with a hyperonic core has been done by Shternin and  Vidaña \cite{ShterninVidana2021}, where they calculate the thermal conductivity, the shear viscosity, and the momentum transfer rates for $np\Sigma^{-}\Lambda eµ$ composition of dense matter in $\beta$–equilibrium based on baryon interactions treated within the framework of the non-relativistic many-body Brueckner-Hartree-Fock theory \cite{RevModPhys.39.719}. Crucially, in their work, they also take into account the in-medium many-body effects on the transport coefficients while treating all the species in their system as non-superfluid. In \cite{Ofengeim_2019}, Ofengeim et al. calculated the bulk viscosity of non-paired $npeµ\Lambda \Xi^-$ matter using the meson-exchange weak interaction.

In the present work, we study the thermal conductivity of a $n p \Lambda$ system in the dense neutron star cores. Our transport calculations incorporate in-medium effects by replacing the bare masses of the constituent particles with their corresponding density-dependent effective masses. Furthermore, the core matter is assumed to be in a non-superfluid and non-superconducting state. Our aim is to qualitatively understand the effects on thermal conductivity of neutrons in the core of neutron stars (in a $\beta$-equilibrium) in the presence of $\Lambda$ hyperons. Thus, by extending the theoretical framework developed for neutrons to include $\Lambda$ baryons, we aim to quantify their contribution to heat transport under realistic stellar conditions. We here use the general formalism developed by Flowers and Itoh \cite{FlowersItoh1976,Flowers1979}, and Yakovlev et al. \cite{1995NuPhA.582..697G, Baiko_2001}, in which, using the variational method, the exact linearized Boltzmann kinetic equations are solved. The thermal conductivity coefficients then calculated are variational solutions. Note that it is also possible to calculate the exact solutions $C_{\text{var}}$ \cite{brooker1968transport, SYKES19701, jensen1968exact, PhysRevLett.21.876}, and they are related to the variational solutions by a correction factor \cite{RezzollaEtAl2017} (see Section 2). Apart from the variational approach, which is traditionally most widely used, there are several other existing frameworks which can be used to calculate the transport coefficients in neutron stars core \cite{Kolomeitsev_2011,PhysRevD.85.103001, Carbone_2011}; a more recent calculation of transport coefficients inside the neutron star core done by Gangopadhyaya et al. \cite{GANGOPADHYAYA2026123367} is based on the relativistic kinetic theory approach using a modified BUU equation with the relaxation time approximation. 

The remainder of the paper is organized as follows: Section \ref{formalism} briefly outlines the theoretical formalism employed. In Section \ref{results}, we present our results and provide a detailed discussion. Finally, our concluding remarks are summarized in Section \ref{conclusion}. Appendix A contains the expressions of integral quantities that appears in Section \ref{formalism}, while Appendix B outlines the generalization of this framework to the systems with more than two heat carrier species.

\section{Formalism}
\label{formalism}

We here consider a neutron star model in which the core at densities between $0.5 n_0$ to $4.5 n_0$ is a dense $np\Lambda e\mu$ matter. One can obtain the total thermal conductivity coefficients of the $np \Lambda e \mu$ system in the dense core of a young neutron star:  $\kappa_{np\Lambda e\mu} = \kappa_{np \Lambda} + \kappa_{e \mu}  \approx \kappa_n + \kappa_p + \kappa_{\Lambda} + \kappa_e + \kappa_{\mu}$, where $\kappa_\alpha$ is the thermal conductivity contribution from $\alpha^{th}$ species. For neutron matter, it can been shown \cite{FlowersItoh1976,Flowers1979} that the thermal conduction of the electron-muon subsystem and the $np\Lambda$ subsystem can be treated independently of each other, because the proton fraction is relatively small and the neutrons and lambdas carry no net charge, hence cannot interact with the charged leptons, neither through the electromagnetic nor through the strong interaction.  Transport coefficients of the $e\mu$ subsystem have been studied in sufficient detail by Shternin and Yakovlev \cite{ShterninYakovlev2007}. Therefore, we here consider the $np\Lambda$ matter and study its thermal conductivity. 
\begin{equation}
\kappa_{np\Lambda}\approx\kappa_n +\kappa_\Lambda \quad \{\text{protons}\sim\text{passive scatterers}\}
\label{eq:1}
\end{equation}
The thermal conductivity of protons is small and can be neglected for the practical purposes \cite{Flowers1979}, and therefore we treat protons as passive scatterers throughout the range of densities considered.

To study the thermal conductivity we use a variational Boltzmann kinetic approach along with Landau-Fermi-liquid theory, where the system of multiple species of fermions is treated as a multi-component Fermi-liquid system which can then be studied using the Fermi-liquid theory accordingly (details of the formalism can be found in \cite{Baiko_2001}).


Thermal conductivity coefficients for a heat carrier species in a Fermi-liquid system can be calculated by taking the simplest variational solution for the distribution function of the species and treating deviations that occurred due to the temperature gradient in the region, away from the equilibrium, as being small enough to be linearized. Then, using Fourier's heat law, one can obtain,

\begin{equation}
    {\kappa_\alpha= \frac{\pi^2}{3} \frac{n_\alpha T}{m^*_\alpha} \tau_\alpha} \quad (\alpha=n,\Lambda)
    \label{eq:2}
\end{equation}\\
where $n_\alpha$ is the number density, $m^*_\alpha$ is the effective mass of the quasi-particle of species $\alpha$ (in our case, neutron and lambda), T is the local temperature, and $\tau_\alpha$ is the thermal relaxation time for the collision of the particle of $\alpha^{th}$ species with particles of other species. 

The thermal relaxation time represents the undetermined part of the variational solution; in the simplest solution, they are taken to be independent of particle energies \cite{ziman1960electrons,RezzollaEtAl2017}. 

\begin{figure*}[t]
  \centering
\includegraphics{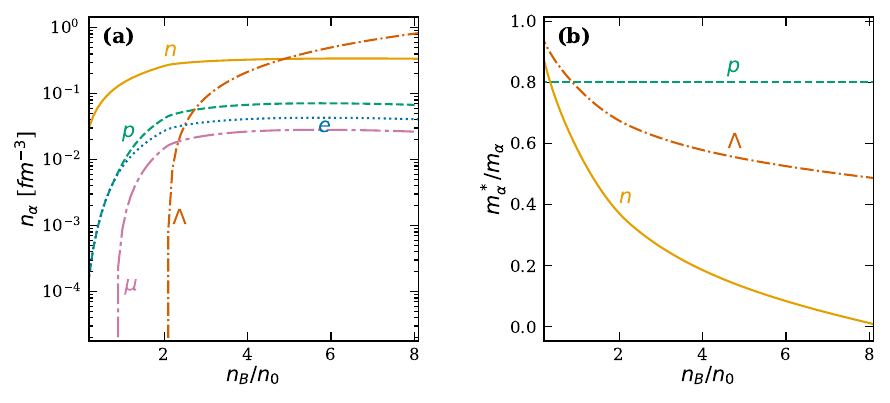}
  \caption{(a) Particle densities and (b) ratio of effective masses to the corresponding bare masses as functions of normalized baryon densities for DDME2 NS EoS \cite{PhysRevC.71.024312}.}
  \label{Figure 1}
 \end{figure*}
 
In our case, we need to solve two linearized Boltzmann kinetic equations, one for neutrons and one for $\Lambda$, to calculate the relaxation times for each component of the Fermi-liquid. We obtain two coupled linear equations in thermal relaxation times.

\begin{equation}
\tau_n(\nu_{nn}+\nu_{np}+\nu_{n\Lambda})+\tau_\Lambda \nu_{n\Lambda}'=1
\label{eq:3}
\end{equation}
\begin{equation}
    \tau_n\nu_{\Lambda n}'+\tau_\Lambda(\nu_{\Lambda \Lambda}+\nu_{\Lambda p}+\nu_{\Lambda n})=1.
    \label{eq:4}
\end{equation}
Here, the coefficients $\nu_{12}$ have the dimensions of frequency and represent the number of collisions per second between particles of species $1$ and $2$. They describe the scattering of a particle of species $1$ $(=n,\Lambda)$ off a particle of species $2$ $(=n,p,\Lambda)$, assumed to be at rest, and are given by

\begin{equation}
    \nu_{12}= \frac{64}{5} \frac{(m_1^*)(m_2^*)^2(k_B T)^2}{\mu_{12}^2 \hbar^3 } {S_{12}}
    \label{eq:5}
\end{equation}
The coefficients $\nu'_{12}$, referred to as the effective collision frequency, are given by
\begin{equation}
    \nu'_{12}= \frac{64}{5} \frac{(m_1^*)^2(m_2^*)(k_B T)^2}{\mu_{12}^2 \hbar^3 } {S'_{12}}
    \label{eq:6}
\end{equation}
for $12=n\Lambda, \Lambda n$. They represent the number of collisions occurring per unit second between particles of species $1$ and $2$ when both the particles are moving, therefore, responsible for the coupled channel behaviour of the system.
Where $\mu_{12}=(m_1 m_2)/(m_1 + m_2)$ is the (bare) reduced mass.

Equations \eqref{eq:3} and \eqref{eq:4} must be solved simultaneously to obtain $\tau_n$ and $\tau_\Lambda$, and consequently the thermal conductivities $\kappa_n$ and $\kappa_\Lambda$. Physically, this reflects the coupled-channel behaviour of a two-component Fermi-liquid system, in which the heat transport by one carrier species, namely $n$ or $\Lambda$, is influenced by the transport of the other species. The quantities $S_{12}$ and $S'_{12}$ have the dimensions of area, each involving a two-dimensional integral that can be evaluated numerically once the required inputs are specified. All eight quantities are discussed in detail and explicitly listed in Eqs.~\eqref{eq:A1}--\eqref{eq:A3} of Appendix~A.

To evaluate $S_{12}$ and $S'_{12}$ we need quantities $|\bar{\mathcal{M}_{\text{12}}}|^2$ the squared absolute of the matrix element for the process $12\rightarrow 1'2'$, averaged over initial and summed over final spin states (here the non-primed quantities refer to the particles before a collision event, while the primed ones refer to the particles after the collision). We can define these quantities in terms of free-space differential cross-section, which are more studied quantities in any physical scattering processes. From the quantum theory of scattering, quantity $|\mathcal{\bar{M}_{\text{12}}}|^2$ is related to free-space differential cross-section in the centre-of-mass (cm) reference frame in which the total momentum of a colliding pair is zero, as (see for example \cite{taylor1972scattering})
\begin{equation}
    \frac{d\sigma_{12}}{d\Omega_{cm}}  = \frac{\mu_{12}^2}{4 \pi^2 \hbar^4} |\bar{\mathcal{M}}_{12}|^2
    \label{eq:7}
\end{equation}

In our calculation, we directly use the fitted form of $S_{nn}$ and $S_{np}$ corresponding to free-space scatterings provided by \cite{Baiko_2001} (they have used the parametrized results from \cite{LiMachleidt1993} for their fit), as they match the experimental $nn$ and $np$ scattering data sufficiently well. It is crucial to note that while this fit matches the experimental data well, they give the best results for any realistic EOS of matter at densities $0.5 n_0 \leq n_B \leq 3n_0$.

 To calculate the free-space differential cross-section for $\Lambda n$ scattering (and, equivalently, for $n\Lambda$ scattering), we use the total cross-section provided by the Nijmegen group \cite{StoksRijken1999} for these scattering channels at laboratory energies below the thresholds for the opening of inelastic channels. We assume the $\Lambda n$ scattering to be isotropic, such that
 \begin{equation}
     \frac{d\sigma_{12}(p_{lab})}{d\Omega_{cm}} \approx \frac {\sigma_{12}(p_{lab})}{4 \pi}
     \label{eq:8}
 \end{equation} \\ 
 These scattering channels are found to be anisotropic \cite{Liu:2026gxr}; nevertheless, we can obtain an order of estimate of the thermal conductivity of our system with the isotropic scattering approximations. For $\Lambda \Lambda$ scattering, in which only the $^1{S}_0$ partial wave channel (which is mainly attractive) contributes significantly \cite{Haidenbauer2016,Polinder2007,Rijken2010}, the isotropic assumption is reasonable.
 Even though practically there are almost no experimental data available on hyperon-hyperon scattering, $\Lambda \Lambda$ models are constrained by the scarce information coming from the $\Lambda \Lambda$ hyper-nuclei (see, for example, \cite{gal2016strangeness} for a systematic review on strangeness in nuclear physics).

\section{Results}
\label{results}

For the numerical evaluation of the transport coefficients, we employ the Density-Dependent Meson-Exchange 2 Equation of State (DDME2 EOS) \cite{PhysRevC.71.024312} within the relativistic mean-field framework with density-dependent meson-baryon couplings. The composition and effective masses obtained from the EOS serve as essential inputs for calculating the $S_{12}$ and $S'_{12}$ integrals, collision frequencies, and subsequently the relaxation times and conductivities.
The left panel of Fig.~\ref{Figure 1} presents the particle number densities of neutrons($n$), protons($p$), $\Lambda$ hyperons, electrons ($e$), and muons($\mu$) as functions of the normalized baryon density $n_B/n_0$, with the nuclear saturation density $n_0 = 0.152\,\mathrm{fm}^{-3}$. $n$ remains the dominant constituent throughout the density range considered, while the $p$ and lepton densities increase gradually under the constraints of charge neutrality and $\beta$-equilibrium. A notable feature is the appearance of $\Lambda$ hyperons at $n_B/n_0 \simeq 2.10$. Beyond this threshold, the $\Lambda$ density rises rapidly and becomes comparable to the $n$ density at higher densities. The onset of $\Lambda$ hyperons marks the transition from purely nucleonic matter to hyperonic matter and significantly modifies the composition of the NS core. The present analysis focuses on the normalized density interval $n_B/n_0 \in [0.5,4.5]$. Following their onset, the $\Lambda$ density increases monotonically with increasing baryon density, indicating their growing importance in the stellar medium. Meanwhile, the proton fraction remains comparatively low across the entire density range. This justifies our treatment of protons as passive scatterers, since their low abundance implies a negligible influence on the overall scattering dynamics compared to the dominant baryonic constituents. 
The right panel of Fig.~\ref{Figure 1} shows the density dependence of the effective masses of $n$, $p$, and $\Lambda$ hyperons. The $n$ effective mass decreases monotonically with increasing density due to the strengthening scalar mean field, reflecting enhanced in-medium effects at supranuclear densities. The $\Lambda$ effective mass exhibits a similar trend but remains larger than the neutron effective mass throughout the density range considered. In contrast, the proton effective mass shows only a weak density dependence and remains nearly constant. These differences arise from the distinct coupling strengths of the baryons to the meson fields and play an important role in determining the scattering rates and transport properties of dense matter.

\begin{figure}[t]
    \centering
    \includegraphics{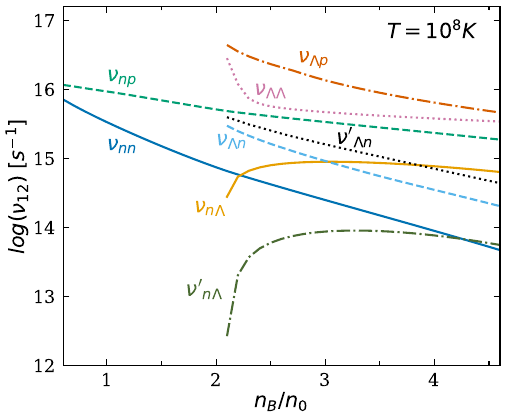}
    \caption{Collision frequencies as functions of normalized baryon densities at temperature $10^8 K$. These are the coefficients that appear in eqs. \eqref{eq:3} and \eqref{eq:4}. Note that the $y$-axis is on a logarithmic scale.}
    \label{Figure 2}
\end{figure}

\begin{figure*}

    \centering
    \includegraphics{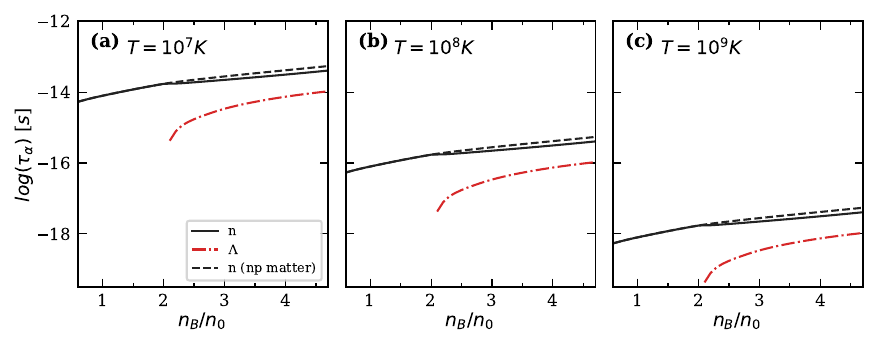}
    \caption{Thermal relaxation times of neutrons (black solid lines) and lambda hyperons (red dashed-dotted lines) in dense $np\Lambda$ matter as functions of normalized baryon densities, plotted for three different temperatures: (a) $10^7 K$, (b) $10^8 K$ and (c) $10^9 K$. The black dashed lines are for the case of $np$ matter.}
    \label{Figure 3}
\end{figure*}

Using the EOS-derived particle fractions and effective masses, along with the free-space scattering differential cross-sections, as inputs, the $S_{12}$ and $S_{12}'$  integrals for the relevant baryon-baryon scattering processes are evaluated (Appendix A). The corresponding collision frequencies are then calculated numerically and subsequently employed in the determination of the relaxation times and thermal conductivities. The variation of collision frequencies with normalized baryon density at $T=10^8$ K has been plotted in the figure. \ref{Figure 2}. With an increase in density, the energy of the particles increases, so the collision probability decreases. Hence, $\nu$ decreases with an increase in density. Or as the density increases, the Fermi momenta of the particles become larger, and the available phase space for scattering is reduced by Pauli blocking, leading to a suppression of collision processes. A change in the slope is observed around \(n_B/n_0 \simeq 2.10\), corresponding to the onset of \(\Lambda\) hyperons, which modifies the matter composition and introduces additional scattering channels. Since the relaxation time is inversely proportional to the collision frequency, the reduction in collision frequencies with density gives rise to the increasing relaxation times.

With this computed $\nu$, $\tau$ for different species have been calculated numerically. Figure \ref{Figure 3} presents the variation of the relaxation time, $\tau_\alpha$, as a function of the normalized baryon density at temperatures $T=10^7$, $10^8$, and $10^9$ K for neutrons and $\Lambda$ hyperons in $\beta$-equilibrated hyperonic matter. For comparison, the neutron relaxation time in purely nucleonic $np$ matter is also shown.  Below the threshold density of $\Lambda$ hyperon appearance, the matter composition consists only of nucleons and leptons, and consequently the neutron relaxation times in hyperonic and nucleonic matter are identical. In all cases, the relaxation times increase with density. Following the onset of $\Lambda$ hyperons, the neutron relaxation time in hyperonic matter becomes smaller than that in pure nucleonic matter due to the opening of additional neutron-hyperon scattering channels, which enhance the collision rate. As the $\Lambda$ population increases with density, the difference between the two neutron curves becomes progressively larger. The $\Lambda$ relaxation time appears only beyond the threshold density and increases monotonically with density. However, it remains smaller than the neutron relaxation time throughout the density range considered. This behavior indicates more frequent collisions involving $\Lambda$ hyperons, arising from both $n\Lambda$ and $\Lambda\Lambda$ scattering processes. A strong temperature dependence is also evident, with the relaxation times decreasing by several orders of magnitude as the temperature increases from $10^7$ K to $10^9$ K. At $T=10^7$ K, the relaxation times are of the order of $10^{-13}$ to $10^{-15}$ s, whereas at $T=10^9$ K they decrease to approximately $10^{-17}$ to $10^{-19}$ s. Despite this quantitative reduction, the overall density dependence and the relative ordering of the neutron and $\Lambda$ relaxation times remain qualitatively unchanged across the temperature range considered.

\begin{figure*}
    \centering
     \includegraphics{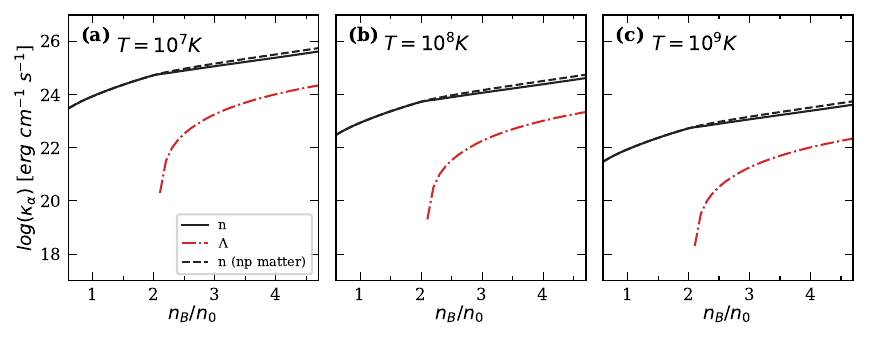}
    \caption{Thermal conductivity of neutrons (black solid lines) and lambda hyperons (red dashed-dotted lines) in dense $np\Lambda$ matter as functions of normalized baryon densities, plotted for three different temperatures: (a) $10^7 K$, (b) $10^8 K$ and (c) $10^9 K$. The black dashed lines are for the case of $np$ matter. In the units of $erg$ $cm^{-1}$ $s^{-1}$.}
    \label{Figure 4}
\end{figure*}
  
Figure \ref{Figure 4} shows the corresponding thermal conductivities of neutrons and $\Lambda$ hyperons as functions of the normalized baryon density at the same temperatures. The neutron thermal conductivity in purely nucleonic matter is included for comparison. To obtain the exact solution from the variational solution, we have multiplied our results with a correction factor $C_{\text{var}}=1.2$ \cite{RezzollaEtAl2017,ShterninVidana2021}. Similar to the relaxation times, the neutron conductivities in hyperonic and nucleonic matter coincide below the $\Lambda$ onset density. Above the threshold, however, the neutron conductivity in hyperonic matter becomes systematically smaller than that in pure nucleonic matter owing to the additional hyperon-induced scattering channels, which reduce the efficiency of heat transport. The suppression becomes more pronounced at higher densities as the $\Lambda$ fraction increases. The $\Lambda$ thermal conductivity emerges only after the appearance of hyperons and rises steadily with density, although it remains two orders of magnitude less below the neutron contribution over the entire density range. This behavior is directly related to the smaller relaxation times of $\Lambda$ hyperons and the stronger scattering experienced by them in the medium.

The thermal conductivity exhibits a marked temperature dependence, decreasing with increasing temperature as a result of enhanced collision frequencies and reduced relaxation times. Consequently, heat transport is more efficient in colder neutron-star interiors and becomes progressively less effective at higher temperatures. Despite the reduction caused by hyperon-induced scattering, the overall conductivity increases with density due to the combined effects of larger particle densities, higher Fermi momenta, and longer relaxation times. \\







 \section{Conclusion}
 \label{conclusion}
 The astrophysical observations indicate the existence of massive neutron stars beyond $2~M_\odot$. Massive neutron stars open the possibility of the appearance of heavier strange baryons inside the core of the star. We estimate the thermal conductivity inside the NS core in presence of $\Lambda$ hyperons. In the work, we investigated the thermal transport properties of dense $np\Lambda$ matter using the DDME2 equation of state. The EOS-derived particle fractions and effective masses were employed to calculate the collision frequencies, relaxation times, and thermal conductivities of neutrons and $\Lambda$ hyperons in $\beta$-equilibrated matter.

The onset of $\Lambda$ hyperons at $n_B/n_0 \simeq 2.10$ introduces additional scattering channels, leading to enhanced collision rates and consequently shorter relaxation times compared to purely nucleonic matter. While the collision frequencies decrease with increasing density due to Pauli blocking and larger Fermi momenta, the relaxation times increase correspondingly. The thermal conductivity is found to increase with density and decrease strongly with temperature, indicating more efficient heat transport in colder neutron-star interiors.

Although the presence of $\Lambda$ hyperons modifies the microscopic scattering processes and reduces the neutron thermal conductivity to some extent, neutrons remain the dominant carriers of heat throughout the density range considered. The contribution of $\Lambda$ hyperons to the total thermal conductivity is comparatively small despite their increasing abundance at higher densities. Therefore, within the DDME2 model \cite{PhysRevC.71.024312}, the inclusion of $\Lambda$ hyperons does not lead to a substantial modification of the overall thermal conductivity of neutron-star matter. It is found that the contribution to thermal conductivity from $\Lambda$ hyperons is not very significant. Even in the presence of $\Lambda$ hyperons, the thermal conductivity contribution from $n$ is not modified substantially. We conclude that while $\Lambda$ hyperons alter the core composition, their net impact on the thermal transport of the $np\Lambda$ system is negligible. This implies that standard nucleon-dominated transport frameworks remain sufficiently accurate for modeling the thermal evolution of dense neutron star cores.
 
\section{Acknowledgement}

The authors express their gratitude to the Nijmegen group \cite{StoksRijken1999} for the public availability of the data that supported this research.

\appendix

\section{$S_{12}$ and $S'_{12}$ Integrals}

\label{app:integrals}

The quantities $S_{12}$ and $S'_{12}$, which have dimensions of area, are given by

\begin{equation}
S_{11}= \frac{\mu_{11}^2}{16 \pi^2 \hbar^4} \int^1_{x'=0} dx' \int_{x=0}^{\sqrt{1-x'^2}} dx \frac{(x^2+x'^2)}{\sqrt{1-x^2-x'^2}} \mid \bar{M}_{11} (x,x') \mid ^2
\label{eq:A1}
\end{equation}

\begin{equation}
S_{12}= \frac{\mu_{12}^2}{16 \pi^2 \hbar^4} \int^{(0.5+x_0)}_{x'=\mid 0.5-x_0 \mid} dx' \int_{x=0}^{a (x')} dx \frac{(1+x^2)}{\sqrt{a^2 (x')- x^2}} \mid \bar{M}_{12} (x,x') \mid ^2
\label{eq:A2}
\end{equation}

\begin{equation}
S'_{12}= \frac{\mu_{12}^2}{16 \pi^2 \hbar^4} \int^{(0.5+x_0)}_{x'=\mid 0.5-x_0 \mid} dx' \int_{x=0}^{a (x')} dx \frac{(0.5+ 2 x_0^2-2 x'^2 -x^2 )}{\sqrt{a^2 (x')- x^2}} \mid \bar{M}_{12} (x,x') \mid ^2
\label{eq:A3}
\end{equation}


where 
\begin{equation}
    a(x')=\frac{\sqrt{x_0^2-(.25+x_0^2-x'^2)}}{x'}
    \label{eq:A4}
\end{equation}

with $x_0 \equiv k_{F2}/{2 k_{F1}}$ being the ratio of Fermi momenta. These quantities can also be defined as the phase-space angular brackets of the corresponding scattering differential cross-sections \cite{Shternin_2020}. Integrations are over the variables $x$ and $x'$, which are defined as \cite{Baiko_2001}

\begin{equation}
    x \equiv \frac{q}{2 k_{F1}}
    \label{eq:A5}
\end{equation}
\begin{equation}
    x' \equiv \frac{q'}{2 k_{F1}}
    \label{eq:A6}
\end{equation}

where the variables $q \equiv |\vec q| =|\vec k'_2 - \vec k_2|$ and $q'\equiv 
    |\vec{q'}|=|\vec k'_2-\vec k_1|$ are relative momenta (note that with this particular choice of $\vec{q}$ and $\vec{q'}$, at the Fermi Surface $\vec{q}\cdot \vec{q'}=0$). The choice of these two independent variables is not unique, and in fact, there are several different choices used in different works (\cite{Baiko_2001}, \cite{ShterninBaldoHaensel2013} and \cite{abrikosov1959theory}). Here we choose to work with $\vec{q}$ and $\vec{q'}$ to follow the notation of Ref.\cite{Baiko_2001}. 

All these integrals, eqs. \eqref{eq:A1} to \eqref{eq:A3},\textemdash there are a total of eight of them corresponding to the eight coefficients in eqs. \eqref{eq:3} and \eqref{eq:4}\textemdash are double integrals with variable limits. To evaluate them numerically, we have changed the integration variables from $x$ to $u$ 
\begin{equation} 
    x=\sqrt{1-x^2}\sin{u}
    \label{eq:A7}
\end{equation} in eqs.\eqref{eq:A1}, and 
\begin{equation}
    x=a(x') \sin{u}
    \label{eq:A8}
\end{equation} in eqs. \eqref{eq:A2} and \eqref{eq:A3}.
When transformed properly, in each of the eight $S_{12}$ and $S'_{12}$ quantities, the integral becomes double integrals with constant limits, and the denominator factors cancel out. These transformed integrals can then be solved using well-known numerical integration techniques. 
We have used Python to evaluate these integrals and calculate the thermal conductivity coefficients.

\section{\texorpdfstring{Generalization to an $N$-heat-carrier-species system}{Generalization to an N-heat-carrier-species system}}
\label{app:generalization}


In this section, we provide the generalization of the formalism to a Fermi-liquid system with more than two components and multiple passive scatterer species. Such a multi-component system, with multiple passive scatter species (for example, see \cite{ShterninVidana2021}), can occur when one considers other hyperons such as $\Sigma^-$ and $\Xi$ in the system, which occur in many neutron star EoS at higher densities (see \cite{Balberg_1999} for details on the occurrence of hyperons in neutron star cores at high densities).
 Consider a dense Fermi-liquid system with N number of carrier species and M number of passive scatterer species (a N-component Fermi-liquid system). In the matrix notation, the N-coupled linear equations for thermal relaxation time of each carrier species are given \cite{RezzollaEtAl2017,ShterninBaldoHaensel2013}

\begin{equation}
\scalebox{0.9}{$
\begin{pmatrix}
\sum_{j=1}^{N+M} \nu_{1j} & \nu'_{12} & \cdots & \nu'_{1N} \\
\nu'_{21} & \sum_{j=1}^{N+M} \nu_{2j} & \cdots & \nu'_{2N} \\
\vdots & \vdots & \ddots & \vdots \\
\nu'_{N1} & \nu'_{N2} & \cdots & \sum_{j=1}^{N+M} \nu_{Nj}
\end{pmatrix}
\begin{pmatrix}
\tau_1 \\
\tau_2 \\
\vdots \\
\tau_N
\end{pmatrix}
=
\begin{pmatrix}
1 \\
1 \\
\vdots \\
1
\end{pmatrix}
$}
\label{eq:B1}
\end{equation}


where the form of coefficients $\nu_{ij}$ and $\nu'_{ij}$ are still given by eqs.\eqref{eq:5} and \eqref{eq:6}, with the quantities $S_{12}$ and $S'_{12}$ given by eqs. \eqref{eq:A1} to \eqref{eq:A3}.

Thus, for an N-component Fermi-liquid system, one needs to solve N coupled linear equations to calculate the thermal conductivity coefficient of each heat carrier species. The total thermal conductivity of the system is given by
\begin{equation}
    \kappa=\sum_{\alpha=1}^N \kappa_\alpha
    \label{eq:B2}
\end{equation}
where $k_\alpha$ is the thermal conductivity of the $\alpha^{th}$ heat carrier species and is given by eq. \eqref{eq:2}.

\bibliography{references}

@article{Baiko_2001,
   title={Thermal conductivity of neutrons in neutron star cores},
   volume={374},
   ISSN={1432-0746},
   url={http://dx.doi.org/10.1051/0004-6361:20010621},
   DOI={10.1051/0004-6361:20010621},
   number={1},
   journal={Astronomy \& Astrophysics},
   publisher={EDP Sciences},
   author={Baiko, D. A. and Haensel, P. and Yakovlev, D. G.},
   year={2001},
   month=jul, pages={151–163} }

@article{Potekhin_2015,
   title={Neutron Stars—Cooling and Transport},
   volume={191},
   ISSN={1572-9672},
   url={http://dx.doi.org/10.1007/s11214-015-0180-9},
   DOI={10.1007/s11214-015-0180-9},
   number={1–4},
   journal={Space Science Reviews},
   publisher={Springer Science and Business Media LLC},
   author={Potekhin, Alexander Y. and Pons, José A. and Page, Dany},
   year={2015},
   month=jul, pages={239–291} }

@article{Gnedin_2001,
    author = {Gnedin, Oleg Y. and Yakovlev, Dmitry G. and Potekhin, Alexander Y.},
    title = {Thermal relaxation in young neutron stars},
    journal = {Monthly Notices of the Royal Astronomical Society},
    volume = {324},
    number = {3},
    pages = {725-736},
    year = {2001},
    month = {06},
    issn = {0035-8711},
    doi = {10.1046/j.1365-8711.2001.04359.x},
    url = {https://doi.org/10.1046/j.1365-8711.2001.04359.x},
    eprint = {https://academic.oup.com/mnras/article-pdf/324/3/725/3394297/324-3-725.pdf},
}

@article{PhysRevB.35.1620,
  title = {Transport properties of a multicomponent Fermi liquid},
  author = {Anderson, R. H. and Pethick, C. J. and Quader, K. F.},
  journal = {Phys. Rev. B},
  volume = {35},
  issue = {4},
  pages = {1620--1629},
  numpages = {0},
  year = {1987},
  month = {Feb},
  publisher = {American Physical Society},
  doi = {10.1103/PhysRevB.35.1620},
  url = {https://link.aps.org/doi/10.1103/PhysRevB.35.1620}
}

@article{PhysRev.185.323,
  title = {Upper and Lower Bounds on Transport Coefficients Arising from a Linearized Boltzmann Equation},
  author = {Jensen, H. H\o{}jgaard and Smith, Henrik and Wilkins, John W.},
  journal = {Phys. Rev.},
  volume = {185},
  issue = {1},
  pages = {323--337},
  numpages = {0},
  year = {1969},
  month = {Sep},
  publisher = {American Physical Society},
  doi = {10.1103/PhysRev.185.323},
  url = {https://link.aps.org/doi/10.1103/PhysRev.185.323}
}

@article{PhysRevD.75.103004,
  title = {Electron-muon heat conduction in neutron star cores via the exchange of transverse plasmons},
  author = {Shternin, P.S. and Yakovlev, D.G.},
  journal = {Phys. Rev. D},
  volume = {75},
  issue = {10},
  pages = {103004},
  numpages = {15},
  year = {2007},
  month = {May},
  publisher = {American Physical Society},
  doi = {10.1103/PhysRevD.75.103004},
  url = {https://link.aps.org/doi/10.1103/PhysRevD.75.103004}
}

@article{Yakovlev_2005,
   title={Neutron star cooling},
   volume={752},
   ISSN={0375-9474},
   url={http://dx.doi.org/10.1016/j.nuclphysa.2005.02.061},
   DOI={10.1016/j.nuclphysa.2005.02.061},
   journal={Nuclear Physics A},
   publisher={Elsevier BV},
   author={Yakovlev, D.G. and Gnedin, O.Y. and Gusakov, M.E. and Kaminker, A.D. and Levenfish, K.P. and Potekhin, A.Y.},
   year={2005},
   month=Apr, pages={590–599} }

@article{Demorest_2010,
   title={A two-solar-mass neutron star measured using Shapiro delay},
   volume={467},
   ISSN={1476-4687},
   url={http://dx.doi.org/10.1038/nature09466},
   DOI={10.1038/nature09466},
   number={7319},
   journal={Nature},
   publisher={Springer Science and Business Media LLC},
   author={Demorest, P. B. and Pennucci, T. and Ransom, S. M. and Roberts, M. S. E. and Hessels, J. W. T.},
   year={2010},
   month=Oct, pages={1081–1083} }

@article{Antoniadis:2013pzd,
    author = "Antoniadis, John and others",
    title = "{A Massive Pulsar in a Compact Relativistic Binary}",
    eprint = "1304.6875",
    archivePrefix = "arXiv",
    primaryClass = "astro-ph.HE",
    doi = "10.1126/science.1233232",
    journal = "Science",
    volume = "340",
    pages = "6131",
    year = "2013"
}

@article{NANOGrav:2019jur,
    author = "Cromartie, H. T. and others",
    collaboration = "NANOGrav",
    title = "{Relativistic Shapiro delay measurements of an extremely massive millisecond pulsar}",
    eprint = "1904.06759",
    archivePrefix = "arXiv",
    primaryClass = "astro-ph.HE",
    doi = "10.1038/s41550-019-0880-2",
    journal = "Nature Astron.",
    volume = "4",
    number = "1",
    pages = "72--76",
    year = "2019"
}

@article{ShterninVidana2021,
author = {P.S. Shternin and Isaac Vida\~{n}a},
title = {Transport coefficients of hyperonic neutron star cores},
journal = {Universe},
year = {2021},
volume = {7},
number = {1},
pages = {1},
doi = {10.3390/universe7010001},
archiveprefix = {arXiv},
eprint = {2106.08474},
primaryclass = {nucl-th}
}

@article{StoksRijken1999,
author = {V. G. J. Stoks and Th. A. Rijken},
title = {Soft-core baryon-baryon potentials for the complete baryon octet},
journal = {Physical Review C},
year = {1999},
volume = {59},
pages = {3009--3030},
doi = {10.1103/PhysRevC.59.3009},
archiveprefix = {arXiv},
eprint = {nucl-th/9901028},
primaryclass = {nucl-th}
}

@article{ShterninBaldoHaensel2013,
author = {P.S. Shternin and M. Baldo and P. Haensel},
title = {Transport coefficients of nuclear matter in neutron star cores},
journal = {Physics Letters B},
year = {2013},
volume = {726},
pages = {850--854},
doi = {10.1016/j.physletb.2013.09.039},
archiveprefix = {arXiv},
eprint = {1311.4278},
primaryclass = {nucl-th}
}

@article{FlowersItoh1976,
author = {E. Flowers and N. Itoh},
title = {Transport properties of dense matter},
journal = {The Astrophysical Journal},
year = {1976},
volume = {206},
pages = {218--231},
doi = {10.1086/154371}
}

@article{LiMachleidt1993,
author = {G. Q. Li and R. Machleidt},
title = {Microscopic calculation of in-medium nucleon-nucleon cross sections},
journal = {Physical Review C},
year = {1993},
volume = {48},
pages = {1702--1712},
doi = {10.1103/PhysRevC.48.1702},
archiveprefix = {arXiv},
eprint = {nucl-th/9307028},
primaryclass = {nucl-th}
}

@incollection{Machleidt_1989,
  author    = {Machleidt, R.},
  title     = {The Meson Theory of Nuclear Forces and Nuclear Structure},
  booktitle = {Advances in Nuclear Physics},
  editor    = {Negele, J. W. and Vogt, Erich},
  volume    = {19},
  pages     = {189--376},
  year      = {1989},
  publisher = {Springer},
  address   = {Boston, MA},
  doi       = {10.1007/978-1-4613-9907-0_2}
}

@article{Brockmann_1990,
  author  = {Brockmann, R. and Machleidt, R.},
  title   = {Relativistic nuclear structure. I. Nuclear matter},
  journal = {Physical Review C},
  year    = {1990},
  volume  = {42},
  number  = {5},
  pages   = {1965--1980},
  doi     = {10.1103/PhysRevC.42.1965}
}

@article{Li_1992,
  author  = {Li, G. Q. and Machleidt, R. and Brockmann, R.},
  title   = {Relativistic calculation of in-medium nucleon-nucleon cross sections},
  journal = {Physical Review C},
  volume   = {45},
  number  = {6},
  pages   = {2782--2794},
  year    = {1992},
  doi     = {10.1103/PhysRevC.45.2782}
}

@article{Machleidt_1987,
  author  = {Machleidt, R. and Holinde, K. and Elster, C.},
  title   = {The Bonn meson-exchange model for the nucleon-nucleon interaction},
  journal = {Physics Reports},
  year    = {1987},
  volume  = {149},
  number  = {1},
  pages   = {1--89},
  doi     = {10.1016/S0370-1573(87)80002-9}
}

@article{Ofengeim_2019,
   title={Bulk viscosity in neutron stars with hyperon cores},
   volume={100},
   ISSN={2470-0029},
   url={http://dx.doi.org/10.1103/PhysRevD.100.103017},
   DOI={10.1103/physrevd.100.103017},
   number={10},
   journal={Physical Review D},
   publisher={American Physical Society (APS)},
   author={Ofengeim, D.D. and Gusakov, M.E. and Haensel, P. and Fortin, M.},
   year={2019},
   month=Nov }

@article{Gusakov_2014,
  author        = {Gusakov, M. E. and Haensel, P. and Kantor, E. M.},
  title         = {Physics of neutron star cores and the $r$-mode instability window},
  journal       = {Monthly Notices of the Royal Astronomical Society},
  year          = {2014},
  volume        = {439},
  number        = {1},
  pages         = {318--333},
  doi           = {10.1093/mnras/stu011},
  eprint        = {1401.2827},
  archiveprefix = {arXiv},
  primaryclass  = {astro-ph.SR}
}

@article{Raduta_2018,
  author        = {Raduta, A. R. and Sedrakian, A. and Weber, F.},
  title         = {Quantum nucleonic core with a hyperonic mantle in neutron stars},
  journal       = {Monthly Notices of the Royal Astronomical Society},
  year          = {2018},
  volume        = {475},
  number        = {4},
  pages         = {4347--4356},
  doi           = {10.1093/mnras/stx3319},
  eprint        = {1712.00584},
  archiveprefix = {arXiv},
  primaryclass  = {astro-ph.HE}
}

@article{Negreiros_2018,
  author        = {Negreiros, R. and Tolos, L. and Centelles, M. and Ramos, A. and Dexheimer, V.},
  title         = {Cooling of Small and Massive Hyperonic Stars},
  journal       = {The Astrophysical Journal},
  year          = {2018},
  volume        = {863},
  number        = {1},
  pages         = {104},
  doi           = {10.3847/1538-4357/aad2d8},
  eprint        = {1804.00334},
  archiveprefix = {arXiv},
  primaryclass  = {astro-ph.HE}
}

@article{PhysRevC.95.025806,
  title = {Lower limit on the heat capacity of the neutron star core},
  author = {Cumming, Andrew and Brown, Edward F. and Fattoyev, Farrukh J. and Horowitz, C. J. and Page, Dany and Reddy, Sanjay},
  journal = {Phys. Rev. C},
  volume = {95},
  issue = {2},
  pages = {025806},
  numpages = {14},
  year = {2017},
  month = {Feb},
  publisher = {American Physical Society},
  doi = {10.1103/PhysRevC.95.025806},
  url = {https://link.aps.org/doi/10.1103/PhysRevC.95.025806}
}

@misc{page2012thermaltransportpropertiesneutron,
      title={Thermal and transport properties of the neutron star inner crust}, 
      author={Dany Page and Sanjay Reddy},
      year={2012},
      eprint={1201.5602},
      archivePrefix={arXiv},
      primaryClass={nucl-th},
      url={https://arxiv.org/abs/1201.5602}, 
}

@article{Burrello2015,
  title = {Heat capacity of the neutron star inner crust within an extended nuclear statistical equilibrium model},
  author = {Burrello, S. and Gulminelli, F. and Aymard, F. and Colonna, M. and Matera, F.},
  journal = {Physical Review C},
  volume = {92},
  issue = {5},
  pages = {055804},
  numpages = {15},
  year = {2015},
  month = {Nov},
  publisher = {American Physical Society},
  doi = {10.1103/PhysRevC.92.055804},
  url = {https://link.aps.org/doi/10.1103/PhysRevC.92.055804}
}

@inproceedings{RezzollaEtAl2017,
author = {Luciano Rezzolla and Pierre Pizzochero and David I. Jones and Nanda Rea and Isaac Vida\~{n}a},
title = {The physics and astrophysics of neutron stars},
booktitle = {Proceedings of the EDP Sciences Conference series on Neutron Stars and Pulsars},
year = {2017},
pages = {1--300},
note = {Lecture notes volume; see editors' volume on neutron star physics},
doi = {10.1051/eas/17xx.xxx}
}

@article{gal2016strangeness,
  title={Strangeness in nuclear physics},
  author={Gal, A and Hungerford, EV and Millener, DJ},
  journal={Reviews of Modern Physics},
  volume={88},
  number={3},
  pages={035004},
  year={2016},
  publisher={APS}
}

@article{brooker1968transport,
  title={Transport properties of a Fermi liquid},
  author={Brooker, GA and Sykes, J},
  journal={Physical Review Letters},
  volume={21},
  number={5},
  pages={279},
  year={1968},
  publisher={APS}
}

@article{SYKES19701,
title = {The transport coefficients of a fermi liquid},
journal = {Annals of Physics},
volume = {56},
number = {1},
pages = {1-39},
year = {1970},
issn = {0003-4916},
doi = {https://doi.org/10.1016/0003-4916(70)90002-3},
url = {https://www.sciencedirect.com/science/article/pii/0003491670900023},
author = {J Sykes and G.A Brooker}
}

@article{jensen1968exact,
  title={Exact transport coefficients for a Fermi liquid},
  author={Jensen, H H{\o}jgaard and Smith, Henrik and Wilkins, John W},
  journal={Physics Letters A},
  volume={27},
  number={8},
  pages={532--533},
  year={1968},
  publisher={Elsevier}
}

@article{ShterninYakovlev2007,
author = {P.S. Shternin and D. G. Yakovlev},
title = {Thermal conductivity of electrons and muons in neutron star cores},
journal = {Physical Review D},
year = {2007},
volume = {75},
pages = {103004},
doi = {10.1103/PhysRevD.75.103004}
}

@article{GlendenningHyperons,
author = {Norman K. Glendenning},
title = {Predictions of hyperons in neutron star cores},
journal = {Nuclear Physics A},
year = {1997},
volume = {625},
pages = {573--581},
doi = {10.1016/S0375-9474(97)00272-5}
}

@article{abrikosov1959theory,
  title={The theory of a fermi liquid (the properties of liquid 3He at low temperatures)},
  author={Abrikosov, AA and Khalatnikov, IM},
  journal={Reports on Progress in Physics},
  volume={22},
  number={1},
  pages={329--367},
  year={1959}
}

@article{Miwa:2022coz,
    author = "Miwa, Koji and others",
    title = "{Recent progress and future prospects of hyperon nucleon scattering experiment}",
    doi = "10.1051/epjconf/202227104001",
    journal = "EPJ Web Conf.",
    volume = "271",
    pages = "04001",
    year = "2022"
}

@misc{miao2024hyperonnucleusnucleonscatteringbesiii,
      title={Hyperon-Nucleus/Nucleon Scattering at BESIII}, 
      author={Han Miao and Jianyu Zhang},
      year={2024},
      eprint={2405.19892},
      archivePrefix={arXiv},
      primaryClass={nucl-ex},
      url={https://arxiv.org/abs/2405.19892}, 
}

@article{Shternin_2020,
   title={Transport coefficients of nucleon neutron star cores for various nuclear interactions within the Brueckner-Hartree-Fock approach},
   volume={102},
   ISSN={2470-0029},
   url={http://dx.doi.org/10.1103/PhysRevD.102.063010},
   DOI={10.1103/physrevd.102.063010},
   number={6},
   journal={Physical Review D},
   publisher={American Physical Society (APS)},
   author={Shternin, P.S. and Baldo, M.},
   year={2020},
   month=sep }

@article{PhysRevC.71.024312,
  title = {New relativistic mean-field interaction with density-dependent meson-nucleon couplings},
  author = {Lalazissis, G. A. and Nik\ifmmode \check{s}\else \v{s}\fi{}i\ifmmode \acute{c}\else \'{c}\fi{}, T. and Vretenar, D. and Ring, P.},
  journal = {Phys. Rev. C},
  volume = {71},
  issue = {2},
  pages = {024312},
  numpages = {10},
  year = {2005},
  month = {Feb},
  publisher = {American Physical Society},
  doi = {10.1103/PhysRevC.71.024312},
  url = {https://link.aps.org/doi/10.1103/PhysRevC.71.024312}
}

@article{Banik_2014,
   title={NEW HYPERON EQUATIONS OF STATE FOR SUPERNOVAE AND NEUTRON STARS IN DENSITY-DEPENDENT HADRON FIELD THEORY},
   volume={214},
   ISSN={1538-4365},
   url={http://dx.doi.org/10.1088/0067-0049/214/2/22},
   DOI={10.1088/0067-0049/214/2/22},
   number={2},
   journal={The Astrophysical Journal Supplement Series},
   publisher={American Astronomical Society},
   author={Banik, Sarmistha and Hempel, Matthias and Bandyopadhyay, Debades},
   year={2014},
   month=sep, pages={22} }

@ARTICLE{Yakovlev_2004,
       author = {{Yakovlev}, D.~G. and {Gnedin}, O.~Y. and {Kaminker}, A.~D. and {Levenfish}, K.~P. and {Potekhin}, A.~Y.},
        title = "{Neutron star cooling: theoretical aspects and observational constraints}",
      journal = {Advances in Space Research},
         year = 2004,
        month = jan,
       volume = {33},
       number = {4},
        pages = {523-530},
          doi = {10.1016/j.asr.2003.07.020},
archivePrefix = {arXiv},
       eprint = {astro-ph/0306143},
 primaryClass = {astro-ph},
       adsurl = {https://ui.adsabs.harvard.edu/abs/2004AdSpR..33..523Y}
}

@article{Tsuruta_2009,
doi = {10.1088/0004-637X/691/1/621},
url = {https://doi.org/10.1088/0004-637X/691/1/621},
year = {2009},
month = {jan},
publisher = {The American Astronomical Society},
volume = {691},
number = {1},
pages = {621},
author = {Tsuruta, S. and Sadino, J. and Kobelski, A. and Teter, M. A. and Liebmann, A. C. and Takatsuka, T. and Nomoto, K. and Umeda, H.},
title = {THERMAL EVOLUTION OF HYPERON-MIXED NEUTRON STARS},
journal = {The Astrophysical Journal}
}

@article{Yakovlev_2004c,
   title={Neutron star cooling: theoretical aspects and observational constraints},
   volume={33},
   ISSN={0273-1177},
   url={http://dx.doi.org/10.1016/j.asr.2003.07.020},
   DOI={10.1016/j.asr.2003.07.020},
   number={4},
   journal={Advances in Space Research},
   publisher={Elsevier BV},
   author={Yakovlev, D.G and Gnedin, O.Y and Kaminker, A.D and Levenfish, K.P and Potekhin, A.Y},
   year={2004},
   pages={523–530} }

@article{RevModPhys.39.719,
  title = {Elements of the Brueckner-Goldstone Theory of Nuclear Matter},
  author = {DAY, B. D.},
  journal = {Rev. Mod. Phys.},
  volume = {39},
  issue = {4},
  pages = {719--744},
  numpages = {0},
  year = {1967},
  month = {Oct},
  publisher = {American Physical Society},
  doi = {10.1103/RevModPhys.39.719},
  url = {https://link.aps.org/doi/10.1103/RevModPhys.39.719}
}

@article{Flowers1979,
  title = {Transport properties of dense matter. II},
  author = {Flowers, Elliott and Itoh, Naoki},
  journal = {The Astrophysical Journal},
  volume = {230},
  pages = {847--858},
  year = {1979},
  publisher = {IOP Publishing},
  doi = {10.1086/156209}
}

@article{Balberg_1999,
doi = {10.1086/313196},
url = {https://doi.org/10.1086/313196},
year = {1999},
month = {apr},
publisher = {},
volume = {121},
number = {2},
pages = {515},
author = {Balberg, Shmuel and Lichtenstadt, Itamar and Cook, Gregory B.},
title = {Roles of Hyperons in Neutron Stars},
journal = {The Astrophysical Journal Supplement Series}
}

@book{ziman1960electrons,
  title={Electrons and Phonons: The Theory of Transport Phenomena in Solids},
  author={Ziman, John Michael},
  year={1960},
  publisher={Clarendon Press},
  address={Oxford}
}

@ARTICLE{1995NuPhA.582..697G,
       author = {{Gnedin}, O.~Y. and {Yakovlev}, D.~G.},
        title = "{Thermal conductivity of electrons and muons in neutron star cores}",
      journal = {\nphysa},
         year = 1995,
        month = feb,
       volume = {582},
        pages = {697-716},
          doi = {10.1016/0375-9474(94)00503-F},
       adsurl = {https://ui.adsabs.harvard.edu/abs/1995NuPhA.582..697G}
}

@article{PhysRevLett.21.876,
  title = {Transport Coefficients of a Normal Fermi Liquid at Finite Temperatures},
  author = {Dy, K. S. and Pethick, C. J.},
  journal = {Phys. Rev. Lett.},
  volume = {21},
  issue = {13},
  pages = {876--878},
  numpages = {0},
  year = {1968},
  month = {Sep},
  publisher = {American Physical Society},
  doi = {10.1103/PhysRevLett.21.876},
  url = {https://link.aps.org/doi/10.1103/PhysRevLett.21.876}
}

@article{GANGOPADHYAYA2026123367,
title = {Calculation of the transport coefficients in neutron star},
journal = {Nuclear Physics A},
volume = {1070},
pages = {123367},
year = {2026},
issn = {0375-9474},
doi = {https://doi.org/10.1016/j.nuclphysa.2026.123367},
url = {https://www.sciencedirect.com/science/article/pii/S037594742600045X},
author = {Utsab Gangopadhyaya and Suman Pal and Gargi Chaudhuri}
}

@ARTICLE{Liu:2026gxr,
       author = {{Liu}, Hang and {Liu}, Liuming and {Tan}, Jin-Xin and {Wang}, Wei and {Yan}, Haobo and {Zhu}, Qian-Teng},
        title = "{A Lattice QCD study of $p-\Lambda$ scattering in continuum and chiral limits}",
      journal = {arXiv e-prints},
         year = 2026,
        month = mar,
          eid = {arXiv:2603.05854},
        pages = {arXiv:2603.05854},
          doi = {10.48550/arXiv.2603.05854},
archivePrefix = {arXiv},
       eprint = {2603.05854},
 primaryClass = {hep-lat},
       adsurl = {https://ui.adsabs.harvard.edu/abs/2026arXiv260305854L}
}

@book{taylor1972scattering,
  author    = {John R. Taylor},
  title     = {Scattering Theory: The Quantum Theory of Nonrelativistic Collisions},
  publisher = {Wiley},
  address   = {New York},
  year      = {1972}
}

@article{RevModPhys.64.1133,
  title = {Cooling of neutron stars},
  author = {Pethick, C. J.},
  journal = {Rev. Mod. Phys.},
  volume = {64},
  issue = {4},
  pages = {1133--1140},
  numpages = {0},
  year = {1992},
  month = {Oct},
  publisher = {American Physical Society},
  doi = {10.1103/RevModPhys.64.1133},
  url = {https://link.aps.org/doi/10.1103/RevModPhys.64.1133}
}

@article{Page_2006,
   title={The cooling of compact stars},
   volume={777},
   ISSN={0375-9474},
   url={http://dx.doi.org/10.1016/j.nuclphysa.2005.09.019},
   DOI={10.1016/j.nuclphysa.2005.09.019},
   journal={Nuclear Physics A},
   publisher={Elsevier BV},
   author={Page, Dany and Geppert, Ulrich and Weber, Fridolin},
   year={2006},
   month=Oct, pages={497–530} }

@article{Zdunik_2013,
   title={Maximum mass of neutron stars and strange neutron-star cores},
   volume={551},
   ISSN={1432-0746},
   url={http://dx.doi.org/10.1051/0004-6361/201220697},
   DOI={10.1051/0004-6361/201220697},
   journal={Astronomy \& Astrophysics},
   publisher={EDP Sciences},
   author={Zdunik, J. L. and Haensel, P.},
   year={2013},
   month=Feb, pages={A61} }

@article{Oertel_2016,
   title={Hyperons in neutron stars and supernova cores},
   volume={52},
   ISSN={1434-601X},
   url={http://dx.doi.org/10.1140/epja/i2016-16050-1},
   DOI={10.1140/epja/i2016-16050-1},
   number={3},
   journal={The European Physical Journal A},
   publisher={Springer Science and Business Media LLC},
   author={Oertel, Micaela and Gulminelli, Francesca and Providência, Constança and Raduta, Adriana R.},
   year={2016},
   month=Mar}

@article{Gnedin_1995,
	doi = {10.1016/0375-9474(94)00503-f},
	url = {https://doi.org},
	year = 1995,
	month = {feb},
	publisher = {Elsevier BV},
	volume = {582},
	number = {3-4},
	pages = {697--726},
	author = {O.Y. Gnedin and D.G. Yakovlev},
	title = {Thermal conductivity of electrons and muons in neutron star cores},
	journal = {Nuclear Physics A}
}

@article{Kolomeitsev_2011,
   title={Superfluid nucleon matter in and out of equilibrium and weak interactions},
   volume={74},
   ISSN={1562-692X},
   url={http://dx.doi.org/10.1134/S1063778811090080},
   DOI={10.1134/s1063778811090080},
   number={9},
   journal={Physics of Atomic Nuclei},
   publisher={Pleiades Publishing Ltd},
   author={Kolomeitsev, E. E. and Voskresensky, D. N.},
   year={2011},
   month=sep, pages={1316–1363} }

@article{PhysRevD.85.103001,
  title = {Shear viscosity in hybrid stars},
  author = {Jaccarino, D. and Plumari, S. and Greco, V. and Lombardo, U. and Santra, A. B.},
  journal = {Phys. Rev. D},
  volume = {85},
  issue = {10},
  pages = {103001},
  numpages = {6},
  year = {2012},
  month = {May},
  publisher = {American Physical Society},
  doi = {10.1103/PhysRevD.85.103001},
  url = {https://link.aps.org/doi/10.1103/PhysRevD.85.103001}
}

@article{Carbone_2011,
  title     = {Transport properties of {$\beta$}-stable nuclear matter},
  author    = {Carbone, A and Benhar, O},
  journal   = {Journal of Physics: Conference Series},
  volume    = {336},
  number    = {1},
  pages     = {012015},
  year      = {2011},
  publisher = {IOP Publishing},
  doi       = {10.1088/1742-6596/336/1/012015},
  url       = {https://doi.org}
}

@article{Logoteta2021,
  title = {Hyperons in Neutron Stars},
  author = {Logoteta, Domenico},
  journal = {Universe},
  volume = {7},
  year = {2021},
  number = {11},
  article-number = {408},
  url = {https://www.mdpi.com/2218-1997/7/11/408},
  doi = {10.3390/universe7110408},
  issn = {2218-1997}
}

@article{Glendenning:1984jr,
    author = "Glendenning, N. K.",
    title = "{Neutron Stars Are Giant Hypernuclei?}",
    doi = "10.1086/163253",
    journal = "Astrophys. J.",
    volume = "293",
    pages = "470--493",
    year = "1985"
}

@article{Bednarek_2012,
   title={Hyperons in neutron-star cores and a 2\,M$_{\odot}$pulsar},
   volume={543},
   ISSN={1432-0746},
   url={http://dx.doi.org/10.1051/0004-6361/201118560},
   DOI={10.1051/0004-6361/201118560},
   journal={Astronomy \& Astrophysics},
   publisher={EDP Sciences},
   author={Bednarek, I. and Haensel, P. and Zdunik, J. L. and Bejger, M. and Mańka, R.},
   year={2012},
   month=jul, pages={A157} }

@article{Yakovlev_2001,
   title={Neutrino emission from neutron stars},
   volume={354},
   ISSN={0370-1573},
   url={http://dx.doi.org/10.1016/S0370-1573(00)00131-9},
   DOI={10.1016/s0370-1573(00)00131-9},
   number={1-2},
   journal={Physics Reports},
   publisher={Elsevier BV},
   author={Yakovlev, D},
   year={2001},
   month=Nov, pages={1–155} }

@article{Sales_2020,
   title={Revisiting the thermal relaxation of neutron stars},
   volume={642},
   ISSN={1432-0746},
   url={http://dx.doi.org/10.1051/0004-6361/202038193},
   DOI={10.1051/0004-6361/202038193},
   journal = {Astronomy \& Astrophysics},
   publisher={EDP Sciences},
   author={Sales, Thiago and Lourenço, Odilon and Dutra, Mariana and Negreiros, Rodrigo},
   year={2020},
   month=Oct, pages={A42} }

@article{Shternin_2022,
   title={Transport coefficients of magnetized neutron star cores},
   volume={58},
   ISSN={1434-601X},
   url={http://dx.doi.org/10.1140/epja/s10050-022-00687-w},
   DOI={10.1140/epja/s10050-022-00687-w},
   number={3},
   journal={The European Physical Journal A},
   publisher={Springer Science and Business Media LLC},
   author={Shternin, Peter and Ofengeim, Dmitry},
   year={2022},
   month=Mar }

@article{Rijken2010,
  author    = {Rijken, Th. A. and Nagels, M. M. and Yamamoto, Y.},
  title     = {Baryon-Baryon Interactions: $\Lambda N$, $\Sigma N$, $\Xi N$, $\Lambda\Lambda$ Interactions --- Extended-Soft-Core Model ESC08},
  journal   = {Progress of Theoretical Physics Supplement},
  volume    = {185},
  pages     = {14--72},
  year      = {2010},
  publisher = {Oxford University Press},
  doi       = {10.1143/PTPS.185.14}
}

@article{Polinder2007,
  author    = {Polinder, H. and Haidenbauer, J. and Mei\ss{}ner, U.-G.},
  title     = {Strangers in nuclei: $YN$ and $YY$ interactions from chiral effective field theory},
  journal   = {Nuclear Physics A},
  volume    = {779},
  pages     = {244--266},
  year      = {2007},
  publisher = {Elsevier},
  doi       = {10.1016/j.nuclphysa.2006.09.002}
}

@article{Haidenbauer2016,
  author    = {Haidenbauer, J. and Mei\ss{}ner, U.-G. and Petschauer, S.},
  title     = {Do different implementations of chiral symmetry imply different properties for the $\Lambda\Lambda$ interaction?},
  journal   = {The European Physical Journal A},
  volume    = {52},
  number    = {1},
  pages     = {15},
  year      = {2016},
  publisher = {Springer},
  doi       = {10.1140/epja/i2016-16015-8}
}

\end{document}